\documentclass[12pt]{elsarticle}

\usepackage{braket}
\usepackage{amsfonts}
\usepackage{color}

\newcommand{\beq}{\begin{eqnarray}}
\newcommand{\eeq}{\end{eqnarray}}
\newcommand{\nn}{\nonumber}

\begin{document}


\title{Deformation quantization of the 
Pais--Uhlenbeck fourth order oscillator}

\author[uaslp]{Jasel Berra--Montiel\corref{cor1}}
\ead{jberra@fc.uaslp.mx}
\author[uaslp,dual]{Alberto Molgado}
\ead{molgado@fc.uaslp.mx}  
\author[uv]{Efra\'\i n Rojas}
\ead{efrojas@uv.mx}

\cortext[cor1]{Corresponding author}

%

\address[uaslp]{Facultad de Ciencias, Universidad Aut\'onoma de San Luis 
Potos\'{\i} \\
Av.~Salvador Nava S/N Zona Universitaria, San 
Luis Potos\'{\i}, SLP, 78290 Mexico}
\address[dual]{Dual CP Institute of High Energy Physics, Mexico}
\address[uv]{Facultad de F\'isica, Universidad Veracruzana, 
Xalapa, Veracruz, 91000 Mexico}

\begin{abstract}
We analyze the quantization of the 
Pais--Uhlenbeck fourth order oscillator within the framework of deformation quantization. Our approach exploits the Noether 
symmetries of the system by proposing 
integrals of motion as the variables to obtain 
a solution to the $\star$-genvalue equation, namely 
the Wigner function. We also obtain, by means of 
a quantum canonical transformation the wave function 
associated to the Schr\"odinger equation of the system.  
We show that unitary evolution of the system 
is guaranteed by means of 
the quantum canonical transformation and via the properties 
of the constructed Wigner function, even in the so called 
equal frequency limit of the model, in agreement with 
recent results.
\end{abstract}





\maketitle

\section{Introduction}
\label{sec:intro}

Whenever one consider curvature terms, for 
example in general relativity or brane inspired models, one is 
faced naturally with field theories described by Lagrangians 
with higher order derivative terms.  
The Pais--Uhlenbeck fourth order linear 
oscillator, originally introduced 
in~\cite{Pais},  is perhaps the simplest example and definitely the best known higher derivative mechanical system and,   
in particular, 
it has served as a toy model to understand several important 
issues related to Ostrogradsky instabilities emerging naturally 
in higher order field 
theories~\cite{gonera,Mannheim,Kosinski,Ghost,Smilga,
pavsic,chen}.  
Recently, the Pais--Uhlenbeck oscillator  has been used as a guide
to study higher order structures associated to 
supersymmetric field theory~\cite{susy}, $PT$-symmetric 
Hamiltonian mechanics~\cite{bender}, and 
geometric models within the scalar field 
cosmology context~\cite{chile}.  
In this sense, it is important to 
mention that naive quantization procedures for 
the Pais--Uhlenbeck model 
has to be enhanced in order to recover unitarity in a physical 
allowed sector. Our main motivation is thus to address 
for the Pais--Uhlenbeck oscillator, within the 
perspective of the deformation quantization formalism, 
the long standing problems associated to the non-unitarity of higher derivative theories.    As we will see, within our formulation
the unitarity is guaranteed straightforwardly, even 
in the equal frequency limit of the model,  by the 
introduction of a well-defined Wigner distribution.

The framework of deformation quantization was introduced in \cite{Flato} as an alternative approach to the problem of quantization. In this formalism one uses, as guidelines, the Dirac quantization rules in order to pass from classical physics to the quantum realm. 
As is to be expected, a consistency requirement for 
such a quantum theory is the existence of a classical limit, that is, a quantum system should reduce to its classical counterpart whenever the limit of $\hbar$, the Planck constant, tends to zero.  
From this perspective, the quantization of a classical system could be seen as a deformation of the 
algebraic structures involved 
in a parameter encoding the quantum nature 
associated to the system ($\hbar$ in our case). Furthermore, the quantization rules require that for any classical observable there is a corresponding quantum observable, and similarly, that the Poisson bracket corresponds to the quantum commutator. All these requirements can be achieved by replacing the usual product of the algebra of smooth functions on the classical phase space with an associative non-commutative product, depending on $\hbar$, such that the resulting commutator is a deformation of the Poisson bracket. We 
refer the reader to~\cite{Groenewold}, where 
results on the explicit construction of maps between classical and quantum observables are explained in detail, 
to Refs.~\cite{Wilde, Fedosov}, where conditions on 
the existence of the star product are exposed, and 
to the reviews~\cite{ZFC,Dito,Waldman} for general aspects
of deformation quantization,  
as well as for more recent 
developments.

Our approach is based on taking advantage of the symmetries 
inherent to the Pais--Uhlenbeck model in order to construct 
the Wigner function that contains the relevant quantum information
of the system.   In this manner, we show that there exists a couple 
of integrals of motion associated to Noether charges, which 
in turn serve as privileged variables in order to find the 
solutions to the $\star$-genvalue equation.   Further, in order 
to obtain the quantum wave functions we consider both, 
classical and quantum canonical transformations.  At a classical 
level we transform, in a standard way, the Pais--Uhlenbeck
system to a simpler model composed of the difference of 
two uncoupled harmonic 
oscillators for which the Wigner function may be also 
obtained.  
We then use the latter Wigner function to obtain a wave equation
which, by means of a quantum canonical transformation, may be 
used in order to obtain the quantum wave function for the 
original Pais--Uhlenbeck system.  The resulting wave function 
is identical to the one obtained in~\cite{Smilga} by a
different reasoning.  We also show that in the equal 
frequency limit of the model the source of the continuous spectrum can be traced out through a linear canonical transformation 
that maps the Pais--Uhlenbeck Hamiltonian to a Hamiltonian 
composed of a discrete spectrum part plus a continuous spectrum
part, contrary to the unequal frequency case.
Besides, we demonstrate that in the equal frequency limit 
the Wigner function is certainly unitary as consequence
of composition of unitary transformations considered through
the quantum canonical transformations.  In this sense, 
our results explicitly manifest the ghost-free feature of the 
Pais--Uhlenbeck model, in complete agreement 
with~\cite{Kosinski,Smilga}.

The article is organized as follows. In 
Section~\ref{sec:Deformation}, we 
include a brief review of deformation quantization in 
order to set our notation and to define some useful 
structures.  We also consider quantum canonical 
transformations as they will be essential in our context
to  obtain the wave functions associated to the 
Pais--Uhlenbeck oscillator.  In Section~\ref{sec:DefPU}, 
we analyze the 
Wigner function for our model in terms of its 
integrals of motion and we identify the quantum wave equation.
Also, in this section we detail the equal frequency limit for 
the Pais--Uhlenbeck oscillator. In Section~\ref{sec:discussion}, 
we include some 
concluding remarks.  Finally, we address some technical 
issues related to the construction of the Pais--Uhlenbeck 
wave function 
in the Appendix.

\section{Deformation quantization}
\label{sec:Deformation}

\subsection{Basic notions}

The origins of deformation quantization were first introduced by H.~Weyl in~\cite{Weyl}.  
The main idea is 
to associate a quantum operator $\hat{W}[f]$ in the Hilbert space 
$L^{2}(\mathbb{R}^{n})$ to every classical observable $f(q,p)$ defined on the phase space  $\mathbb{R}^{2n}$.  This operator is known as the Weyl operator and it is explicitly given by
\beq
f(q,p)\mapsto \hat{W}[f]:=\frac{1}{(2\pi\hbar)^{2n}}\int_{\mathbb{R}^{2n}}d^{n}\eta~ d^{n}\xi\exp\left[\frac{i}{\hbar}(\hat{q}\cdot\eta+\hat{p}\cdot\xi)\right]\tilde{f}(\eta,\xi) \,,
\eeq
\noindent where $\tilde{f}$ is the Fourier transform of $f\in L^{2}(\mathbb{R}^{2n})$ given by
\beq
\tilde{f}(\eta,\xi)=\int_{\mathbb{R}^{2n}}d^{n}q~d^{n}p \exp\left[-\frac{i}{\hbar}(\eta\cdot q+\xi\cdot p)\right]f(q,p) \,,
\eeq
\noindent $\hat{q}$, $\hat{p}$ are operators satisfying the canonical commutations relations, and the integral 
is understood in the weak operator topology \cite{Cattaneo,Esposito}. Later on, Wigner \cite{Wigner} obtained an inverse formula which maps a quantum operator into its symbol, that is, a differential operator with polynomial coefficients defined on classical phase space. This map, known as the Wigner function, 
results a quasi-probability distribution function, explicitly 
written as
\beq\label{eq:WFGeneric}
\rho(q,p)=\frac{1}{(2\pi)^{n}}\int_{\mathbb{R}^{n}}d^{n}y~\psi^{*}
\left(q-\frac{\hbar}{2}y\right)
e^{-iy\cdot p}\psi
\left(q+\frac{\hbar}{2}y\right) \,.
\eeq  
\noindent Among the most important properties of the 
Wigner function 
we would like to point out the following.  In the first place, 
Wigner function is a particular representation of the density matrix which is normalized, bounded and real.  
In 
the second place, for a given quantum wave function $\psi(x)$, Wigner function represents a generating function for all spatial auto-correlations.  Finally, in the classical limit $\hbar\mapsto 0$ Wigner function reduces to a highly localized probability density in the coordinate space~\cite{Zachos}.  

Subsequently, Moyal found an explicit formula for the symbol of a quantum commutator between operators \cite{Moyal},
\beq
(f\star g)(q,p)=f\exp\left(\frac{i\hbar}{2}\overleftarrow{\partial}_{q^{i}}\overrightarrow{\partial}_{p_{i}}-\frac{i\hbar}{2}\overleftarrow{\partial}_{p_{i}}\overrightarrow{\partial}_{q^{i}}\right)g \,. 
\eeq    
This $\star$-product is known as the Moyal product, and will be defined as an involution in~(\ref{eq:involution}). For a two dimensional case ($n=1$) this deformed product reads 
\beq
f\star g=\sum_{k=0}^{\infty}\frac{1}{k!}\left( \frac{i\hbar}{2} \right)^{k}\sum_{m=0}^{k}{k \choose m}(-1)^{m}(\partial_{q}^{k-m}\partial_{p}^{m}f)(\partial_{q}^{m}\partial_{p}^{k-m} g) \,. 
\eeq 
Straightforwardly, one may check that the Moyal product satisfies the following properties: 
\begin{enumerate}
\item  $f\star g=fg+\mathcal{O}(\hbar)$, that is, the 
$\star$-product is a deformation of the usual pointwise product between smooth functions.
\item With respect to the $\star$-product, 
the Weyl operator 
$\hat{W}$ is a homomorphism of algebras,
\beq
\hat{W}:\left(C^{\infty}(\mathbb{R}^{2n}),\star\right)\rightarrow \left(\mathcal{L}(L^{2}(\mathbb{R}^{n})),\circ\right)  \,,
\eeq
where
\beq
\hat{W}[f\star g]:=\hat{W}[f]\circ\hat{W}[g] \,,
\label{eq:involution}
\eeq
between distributional functions defined on the phase space and the space of bounded operators on the square integrable functions $L^{2}(\mathbb{R}^{n})$ under the composition $\circ$ as a product. 
\item The deformed Poisson bracket associated to 
the $\star$-product is given by
\beq
\label{eq:Moyal--Poisson}
\left[f,g \right]_{\star}:=\frac{1}{i\hbar}\left(f\star g-g\star f \right)=\left\lbrace f,g \right\rbrace +\mathcal{O}(\hbar) \,,   
\eeq
where $\left\lbrace \cdot,\cdot\right\rbrace$ denotes the standard Poisson bracket, and $\left[\cdot,\cdot\right]_{\star}$ is known as the Moyal commutator or the Moyal bracket.  
As a consequence of~(\ref{eq:Moyal-Poisson}),
Moyal bracket may be interpreted 
as a Lie bracket on the space of Weyl operators~\cite{Takhtajan}. 
\end{enumerate}

In order to avoid convergence problems, the Moyal product is defined not in the space of smooth functions, $C^{\infty}(\mathbb{R}^{2n})$, but in the extended space, $C^{\infty}(\mathbb{R}^{2n})[[\hbar]]$, which corresponds to the space of formal power series in $\hbar$ with coefficients in $C^{\infty}(\mathbb{R}^{2n})$, or 
alternatively, it can be defined on the space of Schwartz functions  using an integral representation and then extend it to a suitable space of distributions \cite{Garcia,Kammerer}. The interpretation of this product as a non-commutative deformation on the algebra of observables was introduced in~\cite{Flato} by means of Gerstenhaber's algebras.  
In particular,  the existence of a 
$\star$-product for an arbitrary symplectic manifold was
demonstrated in \cite{Fedosov}. Subsequently, as a consequence of the formality theorem, Kontsevich solved the existence and classification problem of star products on a generic finite dimensional Poisson manifold \cite{Kontsevich}. 

\subsection{Time evolution and canonical transformations}
\label{sec:canonicalT}

The expectation values of observables and the time evolution of states can be computed in a similar manner as in classical mechanics, however, the usual pointwise product between functions and the Poisson bracket are replaced with the $\star$-product and the Moyal commutator, respectively. In consequence, 
the expectation value of an observable $A$ in a state $\psi$, is given by the expression
\beq
\left\langle A \right\rangle_{\psi}=\frac{1}{(2\pi\hbar)^{n/2}}\int_{\mathbb{R}^{2n}}d^{n}qd^{n}p\, A(q,p)\star \rho(q,p) \,, 
\eeq   
\noindent where $\rho(x,p)$ is the Wigner function defined in (\ref{eq:WFGeneric}).  In this sense, expectation values of physical observables in deformation quantization correspond to obtaining the trace of an operator with the density matrix, in close analogy to standard probability theory~\cite{Zachos}.

The dynamical equation of the quantum distribution $\rho(x,p)$, results in the counterpart of Liouville's theorem in classical mechanics describing the time evolution of a classical distribution function, and is given by the formula
\beq
\frac{\partial \rho(q,p)}{\partial t}=\frac{1}{i\hbar}\left[H,\rho\right]_{\star}=\frac{H\star\rho-\rho\star H}{i\hbar} \,, 
\eeq 
\noindent where $H$ is a distinguished real function from the algebra of observables, namely, the Hamiltonian. Being time-dependent, this evolution equation, also known as Moyal's equation, does not completely 
determine the Wigner function for a system \cite{Takahashi}. Then, just as in the conventional formulation of quantum mechanics, a systematic solution 
may be inferred from the spectrum of the stationary problem. Static Wigner functions obey a more suggestive $\star$-genvalue equation, inducing Bopp
shifts~\cite{ZFC} 
\beq\label{genvalue}
H(q,p)\star\rho(q,p)
& = & 
\rho(q,p)\star H(q,p)=H\left(q+\frac{i\hbar}{2}\partial_{p},p-\frac{i\hbar}{2}\partial_{q} \right)
\rho(q,p) \nn\\
& = & 
E\rho(q,p)
\,,
\eeq
\noindent where $E$ corresponds to the energy eigenvalue associated to the Hamiltonian, leading to the spectral properties of the Wigner function as a quantum distribution function~\cite{ZFC}. These quantum properties are related to the fact that Wigner function is not positive semi-definite, 
allowing in principle 
negative values for 
certain areas of the phase space. This 
counter intuitive negative-probability aspect has been 
speculated as a way to detect  
quantum interference which may be  
measured and reconstructed indirectly in the laboratory \cite{Bertet}. 

As within the framework of deformation quantization 
the observables are represented by smooth functions defined on  phase space, and thus they transform classically, the outcome of a canonical transformation on the quantum $\star$-genvalue equations results in an appropriate transformation of the Wigner function \cite{Fairlie}.

\noindent A general transformation of the phase space coordinates is defined as a smooth bijective map $T:\mathbb{R}^{2n}\ni (q,p)\rightarrow (Q,P)\in \mathbb{R}^{2n}$, which transforms every observable $A\in C^{\infty}(\mathbb{R}^{2n})$ by
\beq\label{trans}
A'=A\circ T.
\eeq

\noindent Further, the star product in the new variables 
$\star'$ satisfies the natural condition \cite{Blaszak}
\beq
(f\star g)\circ T=(f\circ T)\star'(g\circ T), \;\;\; f,g\in C^{\infty}(\mathbb{R}^{2n}) \,,
\eeq 
in such a way that the $\star'$-product is given by
\beq
f\star' g=f\exp\left(\frac{i\hbar}{2}\overleftarrow{D}_{Q^{i}}\overrightarrow{D}_{P_{i}}-\frac{i\hbar}{2}\overleftarrow{D}_{P_{i}}\overrightarrow{D}_{Q^{i}} \right)g \,, 
\eeq
\noindent where the vector fields $D_{Q^{i}}$, $D_{P_{i}}$ correspond to the transformed derivations $\partial_{q^{i}}$, $\partial_{p_{i}}$ according to the rule stated in equation (\ref{trans}).

Among the phase space coordinate maps there are some which play a special role in classical mechanics, 
namely the canonical transformations or symplectomorphisms, that is, those maps which preserve the 
symplectic form, and thus the Poisson bracket structure
\beq
\left\lbrace q^{i},p_{j}\right\rbrace =\left\lbrace Q^{i},P_{j}\right\rbrace =\delta^{i}_{j} \,. 
\eeq
\noindent In phase space quantum mechanics, a quantum canonical transformation is a transformation $T$ such that it preserves the form of the corresponding deformed Poisson bracket, namely, the Moyal's bracket
\beq
\left[ q^{i},p_{j}\right]_{\star}=\left[ Q^{i},P_{j}\right]_{\star'}=\delta^{i}_{j} \,,  
\eeq
\noindent where $\left[\cdot,\cdot \right]_{\star'} $ denotes the Moyal's bracket transformed by $T$ to the new coordinate system \cite{Blaszak,Moshinsky},
\beq
\left[f,g \right]_{\star'}=\left[f\circ T^{-1},g\circ T^{-1} \right]_{\star}\circ T \,.  
\eeq
The transformation of coordinates $T$ induces a unitary operator on the Hilbert space $\hat{U}_{T}:L^{2}(\mathbb{R}^{n})\rightarrow L^{2}(\mathbb{R}^{n})$, transforming vector states and observables to the new coordinate system in such a way that 
for any quantum observable $\hat{A}$ the transformation reads
\beq\label{transobs}
\hat{A}'(\hat{Q},\hat{P})=\hat{U}_{T}\hat{A}(\hat{q},\hat{p})\hat{U}_{T}^{-1}  \,.
\eeq   
\noindent This expression allows us to derive the form of the operator $\hat{U}_{T}$, which corresponds to the celebrated Dirac's quantum transformation \cite{Dirac} 
\beq\label{QCT}
(\hat{U}_{T}\psi)(Q)=\frac{1}{(2\pi\hbar)^{n/2}}\int d^{n}Q\, \sqrt{\left|\frac{\partial^{2}F}{\partial q\partial Q}(q,Q)\right|}e^{\frac{i}{\hbar}F(q,Q)}\psi(q) \,.
\eeq
being $F(q,Q)$ the generator of the 
classical canonical transformation such that 
\beq
p  =  \frac{\partial F(q,Q)}{\partial q} \,,
\hspace{5ex} 
P  =  -\frac{\partial F(q,Q)}{\partial Q}  \,.
\eeq
\noindent In this expression, the implementation of the quantum canonical transformation on state eigenfunctions is realized through a generalization of a Fourier-type transformation 
containing $\hbar$-corrections with respect to the classical canonical counterpart, thus encompassing the quantum nature of the operator. In addition, it can be demonstrated that $\hat{U}_{T}$ is indeed a unitary operator on the Hilbert space $L^{2}(\mathbb{R}^{n})$~\cite{Fairlie,Blaszak}.

\section{Deformation quantization of the Pais--Uhlenbeck Oscillator}
\label{sec:DefPU}

In this section, we perform a deformation quantization of the Pais--Uhlenbeck oscillator by solving the corresponding $\star$-genvalue equations and, in order to obtain the associated wave functions, 
we make use of the theory of quantum canonical transformations discussed in the previous section.

\subsection{Wigner function}

In order to study the quantum dynamics of the system, we will start by considering the Lagrangian 
\beq
L(q,\dot{q},\ddot{q})=\frac{1}{2}\left[ \ddot{q}^{2}-(\Omega_{1}^{2}+\Omega_{2}^{2})\dot{q}^{2}+\Omega_{1}^{2}\Omega_{2}^{2}
q^2 \right] \,, 
\eeq
\noindent where $q$ is a real-valued function of time, $\dot{q}$ and 
$\ddot{q}$ its first and second time-derivatives, respectively, and the parameters $\Omega_{1}$ and $\Omega_{2}$ correspond to a pair of frequencies which are taken real and positive. More explicitly, this model is characterized by the following fourth-order differential equation of motion
\beq\label{eom}
\frac{d^{4}q}{dt^{4}}+(\Omega_{1}^{2}+\Omega_{2}^{2})\frac{d^{2}q}{dt^{2}}+\Omega_{1}^{2}\Omega_{2}^{2}q=0 \,,
\eeq
\noindent which can be derived directly from the Lagrangian. The canonical Hamiltonian can be obtained by means of the Ostrogradsky's method \cite{Ostrogradsky}. To implement this method, the phase space involves, 
in addition to the canonical coordinates $(q,p_{q})$, 
an extra canonical pair of variables, namely $x:=\dot{q}$ with corresponding canonical momentum $p_{x}$ \cite{Ghost,Smilga,chen}. 
The momenta are defined as $p_x := \partial L/\partial \ddot{q}$ 
and $p_q  :=  \partial L/\partial \dot{q} - d/dt \left( 
\partial L/\partial \ddot{q} \right)$, and in our case they 
explicitly read
\beq
\nn
p_x & = & \ddot{q}  \,,\\
p_q & = & -(\Omega_1^2+\Omega_2^2)\dot{q} -  \frac{d^3q}{dt^3} \,,
\eeq
respectively, while the canonical Hamiltonian is obtained 
through the Legendre transformation 
$H(q,p_q,x,p_x)=p_q\dot{q}+p_x\dot{x}-L(q,\dot{q},\ddot{q})$,
and thus
\beq\label{Ham}
H(q,p_q,x,p_x)=p_{q}x+\frac{p_{x}^{2}}{2}+\frac{\left(\Omega_{1}^{2}+\Omega_{2}^{2} \right)x^{2} }{2}-\frac{\Omega_{1}^{2}\Omega_{2}^{2}q^{2}}{2}  \,.
\eeq
For this theory, it is natural to define the generalized Poisson bracket 
\beq\label{Poisson}
\left\lbrace A,B\right\rbrace :=\frac{\partial A}{\partial q}\frac{\partial B}{\partial p_{q}}-\frac{\partial A}{\partial p_{q}}\frac{\partial B}{\partial q}+\frac{\partial A}{\partial x}\frac{\partial B}{\partial p_{x}}-\frac{\partial A}{\partial p_{x}}\frac{\partial B}{\partial x} \,, 
\eeq
\noindent in order to obtain canonical relations
\beq
\left\lbrace q,p_{q}\right\rbrace =1= \left\lbrace x,p_{x} \right\rbrace \,,  
\eeq
while the rest of the brackets among the phase space variables 
are vanishing.
\noindent This Poisson structure enable us to write the canonical Hamilton equations of motion which, by making 
the identification $x=\dot{q}$, straightforwardly lead to equation~(\ref{eom}). Therefore, the Hamiltonian 
$H$, obtained in~(\ref{Ham}), describes the dynamics in phase space for the fourth order Pais--Uhlenbeck model, and thus this Hamiltonian corresponds to the one which is to be quantized, using 
in our case, the 
techniques related to deformation quantization
as reviewed above. 

Our goal now is to calculate the Wigner function.  
As we will see below, the $\star$-genvalue problem~(\ref{genvalue}) can be solved directly without first solving the corresponding 
Schr\"odinger equation. 
Considering 
the Pais--Uhlenbeck oscillator with Hamiltonian given by (\ref{Ham}), the $\star$-genvalue equation (\ref{genvalue}) for the Wigner 
function $\rho(q,p_{q},x,p_{x})$ 
explicitly reads
\begin{eqnarray}\label{MoyalPais}
\left[ 
\left(p_{q}-\frac{i\hbar}{2}\partial_{q} \right)\left(x+\frac{i\hbar}{2}\partial_{p_{x}}\right) 
+ \left( \frac{\Omega_{1}^{2}
+\Omega_{2}^{2}}{2}\right)\left(x+\frac{i\hbar}{2}\partial_{p_{x}} \right)^{2} \right. & &  
\nn\\ 
+\frac{1}{2}\left(p_{x}-\frac{i\hbar}{2}\partial_{x} \right)^{2} - \left.\frac{\Omega_{1}^{2}\Omega_{2}^{2}}{2}\left( q+\frac{i\hbar}{2}\partial_{p_{q}}\right)^{2}\right]\rho
& = &  
E\rho \,,
\end{eqnarray}      
\noindent where we have used the associative $\star$-product 
\beq
\star:=\exp\left[ \frac{i\hbar}{2}\left(\overleftarrow{\partial_{q}}\overrightarrow{\partial_{p_{q}}}-
\overleftarrow{\partial_{p_{q}}}\overrightarrow{\partial_{q}}+\overleftarrow{\partial_{x}}
\overrightarrow{\partial_{p_{x}}}- \overleftarrow{\partial_{p_{x}}}\overrightarrow{\partial_{x}} \right)   \right], 
\eeq
\noindent which is naturally associated to 
the canonical Poisson structure~(\ref{Poisson}) 
of the Pais--Uhlenbeck oscillator. The eigenvalue equation (\ref{MoyalPais}) 
may be decomposed into two partial differential equations corresponding to the real and imaginary parts of the equation, respectively. These equations explicitly read
\beq \label{MP2}
\left[ p_{q}x+\frac{p_{x}^{2}}{2}+\frac{\hbar^{2}}{4}\left( \partial^{}_{q}\partial_{p_{x}}-\frac{1}{2}\partial^{2}_{x}\right)-\frac{\Omega_{1}^{2}+\Omega_{2}^{2}}{2}\left(x^{2}-\frac{\hbar^{2}}{4}\partial^{2}_{p_{x}} \right) \right. \nn\\
\left. -\frac{\Omega_{1}^{2}\Omega_{2}^{2}}{2}\left( q^{2}-\frac{\hbar^{2}}{4}\partial^{2}_{p_{q}}\right)-E\right]\rho(q,p_{q},x,p_{x})=0 \,,
\eeq
for the real part, while 
\beq
\left[p_{q}\partial_{p_{x}}-x\partial_{q}-p_{x}\partial_{x}+\left(\Omega_{1}^{2}+\Omega_{2}^{2} \right)x\partial_{p_{x}}-\Omega_{1}^{2}\Omega_{2}^{2}q\partial_{p_{q}}  \right]\rho(q,p_{q},x,p_{x})=0 \,, 
\eeq
corresponds to the imaginary part.
Although, these equations may seem challenging
at first sight, we can take advantage of the deeply relation between 
Moyal quantization and the canonical structure of the theory in 
order to find solutions to these equations. By using the symmetries of the Hamiltonian, and employing the conserved Noether charges as variables for the Wigner function, one can infer the solution to the $\star$-genvalue problem~\cite{Koikawa}.
Indeed, using Noether theorem, suitable generalized to higher derivative theories, one finds that the Pais--Uhlenbeck oscillator remains invariant under the symmetry~\cite{Bolonek},   
\beq
\label{eq:infinitransform}
q\mapsto q+\varepsilon\left(\frac{d^{3}q}{dt^{3}}\pm(\Omega_{1}^{2}-\Omega_{2}^{2})\frac{dq}{dt}\right) \,, 
\eeq
\noindent which, as a consequence, imply the existence of two global integrals 
of motion
\begin{eqnarray}
\label{eq:J1}
J_{1}&=& \frac{1}{\Omega_{1}^{2}-\Omega_{2}^{2}}\left[ \Omega_{1}^{2}\left(p_{x}+\Omega_{2}^{2}q \right)^{2}+\left(p_{q}+\Omega_{1}^{2}x \right)^{2}  \right] \,, \\
\label{eq:J2} 
J_{2}&=&\frac{1}{\Omega_{1}^{2}-\Omega_{2}^{2}}\left[ \left(p_{q}+\Omega_{2}^{2}x \right)^{2}+\Omega_{2}^{2}\left(p_{x}+\Omega_{1}^{2}q \right)^{2}  \right] \,.
\end{eqnarray}
These integrals of motion are associated to 
infinitesimal transformations of the  
form~(\ref{eq:infinitransform}).  However, when one considers
infinitesimal transformations along the velocity vector field,  
$\dot{q}$, together with infinitesimal time transformations, the integrals of motion are associated to the 
so-called energies for a second order differential Lagrangian.
In particular, one may check that the Pais--Uhlenbeck 
Hamiltonian~(\ref{Ham}) is related to the integrals of motion of 
our interest by the identity $H=(J_1-J_2)/2$.  A general 
statement of Noether theorem and the energies for a second order Lagrangian system is reviewed in~\cite[Appendix A]{CGMR}, while a detailed exposition 
may be found in~\cite{Miron}. 

Thus, using constants of motion~(\ref{eq:J1}) and~(\ref{eq:J2}), the Wigner function given by
\beq \label{WF}
\rho_{nm}(q,p_{q},x,p_{x})=\frac{(-1)^{m+n}}{\pi^{2}\hbar^{2}}e^{-2J_{1}/\hbar\Omega_{1}}e^{-2J_{2}/\hbar\Omega_{2}}
L_{n}\left(\frac{4J_{1}}{\hbar\Omega_{1}}\right)
L_{m}\left(\frac{4J_{2}}{\hbar\Omega_{2}}\right) \,,
\eeq
\noindent corresponds to a solution to the $\star$-genvalue equation (\ref{MoyalPais}). 
Here, the $L$'s 
stand for the Laguerre polynomials defined 
by the Rodrigues formula
\beq
L_{n}(z)=\frac{e^{z}\partial^{n}(e^{-z}z^{n})}{n!} \,.
\eeq

As we will see in the next section, 
Wigner function~(\ref{WF}) 
is intrinsically related to the Wigner function of the harmonic oscillator. Besides, it turns out that the Hamiltonian of the 
Pais--Uhlenbeck oscillator is canonically equivalent to the Hamiltonian corresponding to the difference of two uncoupled harmonic oscillators. This imply, as long as the canonical transformation remains linear \cite{Moshinsky}, that the Wigner function of the Pais--Uhlenbeck is equivalent to the Wigner function of a pair of harmonic oscillators, but written in terms of its own Noether charges.  

Also, after some algebraic manipulations, one may check by substituting the Wigner 
formula~(\ref{WF}) into equation (\ref{MP2}), that the system has 
energy $\star$-genvalues
\beq\label{genvalues}
E_{nm}=\left( n+\frac{1}{2}\right)\Omega_{1}-\left( m+\frac{1}{2}\right)\Omega_{2} \,, 
\hspace{7ex} 
n,m=0,1,2,\ldots \,, 
\eeq
as it is well-known for the Pais--Uhlenbeck
oscillator.

\subsection{Wave function}

In this section we are interested in 
obtaining the wave function associated to the quantum mechanical system of the Pais--Uhlenbeck oscillator within the framework of deformation quantization. To this end, we will use 
the formalism of quantum canonical transformations already discussed in section~\ref{sec:canonicalT}. This can be achieved by following the next strategy. First, by means of an appropriate canonical transformation, 
we will map the Hamiltonian of the Pais--Uhlenbeck oscillator into a Hamiltonian describing a 
pair of uncoupled harmonic oscillators. The second 
step will be to solve the Wigner function corresponding to the new Hamiltonian, and then, through a Fourier transformation, we will 
calculate the wave function related to the harmonic oscillator problem. Finally, we 
will consider a quantum canonical transformation in order to obtain the wave function belonging to the original problem, that is, 
the Pais--Uhlenbeck oscillator. 
The aim of this approach is twofold.  On the one hand, instead of calculating the wave function directly from the Pais--Uhlenbeck Wigner distribution, 
we note that 
the quantum canonical transformation not only 
leads to more manageable integrals, but also it
allows us to compare with previous results found in the literature where the wave function is obtained by solving the Schr\"{o}dinger equation.  On the other hand, the equal frequency limit 
$\Omega_{1}\rightarrow \Omega_{2}$ 
is analyzed via canonical transformations
at both, the classical and the quantum 
levels~\cite{Mannheim,Smilga}.

Consider for definiteness,
the case of unequal frequencies and assume $\Omega_{1}>\Omega_{2}$. The Hamiltonian (\ref{Ham}) can be brought into diagonal form by applying the canonical transformation \cite{Smilga},\begin{eqnarray}
q&=& \frac{1}{\Omega_{1}}\frac{\Omega_{1}X_{2}-P_{1}}{\sqrt{\Omega_{1}^{2}-\Omega_{2}^{2}}} \,, \;\;\;\;\;\;\; x=\frac{\Omega_{1}X_{1}-P_{2}}{\sqrt{\Omega_{1}^{2}-\Omega_{2}^{2}}} \,, \nn\\
p_{x}&=&\frac{\Omega_{1}P_{1}-\Omega_{2}^{2}X_{2}}{\sqrt{\Omega_{1}^{2}-\Omega_{2}^{2}}} \,, \;\;\;\;\;\;\;\; p_{q}=\Omega_{1}\frac{\Omega_{1}P_{2}-\Omega_{2}^{2}X_{1}}{\sqrt{\Omega_{1}^{2}-\Omega_{2}^{2}}} \,,
\end{eqnarray}
\noindent which is realized by the generating function 
\beq \label{generator}
F(q,x,X_{1},X_{2})=\Omega_{1}\gamma qX_{2}+\gamma xX_{1}-\Omega_{1}^{2}qx-\Omega_{1}X_{1}X_{2}
\,,
\eeq 
where $\gamma:=\sqrt{\Omega_{1}^{2}-\Omega_{2}^{2}}$. Using such a canonical transformation, one finds that the Hamiltonian is mapped into the difference of an uncoupled pair of harmonic oscillators
\beq\label{Ham2}
H(X_1,P_1,X_2,P_2)=\frac{P_{1}^{2}+\Omega_{1}^{2}X_{1}^{2}}{2}-\frac{P_{2}^{2}+\Omega_{2}^{2}X_{2}^{2}}{2} \,.
\eeq
Following the $\star$-genvalue equation (\ref{genvalue}) for this Hamiltonian, the resulting equation is
\begin{eqnarray}
&&\left[ \left( P_{1}-\frac{i\hbar}{2}\partial_{X_{1}}\right)^{2}+\Omega_{1}^{2}\left(X_{1}+\frac{i\hbar}{2}\partial_{P_{1}} \right)^{2}-\left( P_{2}-\frac{i\hbar}{2}\partial_{X_{2}}\right)^{2} \right. \nn\\
&&\left.-\Omega_{2}^{2}\left(X_{2}+\frac{i\hbar}{2}\partial_{P_{2}} \right)^{2}-2E \right]\rho(X_{1},X_{2},P_{1},P_{2})=0 \,.
\end{eqnarray}
\noindent By virtue of its imaginary part
\beq
\hbar\left( -P_{1}\partial_{X_{1}}+\Omega_{1}^{2}X_{1}\partial_{P_{1}}+P_{2}\partial_{X_{2}}-\Omega_{2}^{2}X_{2}
\partial_{P_{2}}\right)\rho=0, 
\eeq
\noindent the Wigner function $\rho$ is seen to depend on two variables, $z_{1}=4H_{1}^{\mathrm{osc}}/\hbar\Omega_{1}$ and $z_{2}=4H_{2}^{\mathrm{osc}}/\hbar\Omega_{2}$, where  $H_{1}^{\mathrm{osc}}:=(P_{1}^{2}+\Omega_{1}^{2}X_{1}^{2})/2$ and 
$H_{2}^{\mathrm{osc}}:=(P_{2}^{2}+\Omega_{2}^{2}X_{2}^{2})/2$ correspond to the Noether charges to the Hamiltonian of uncoupled harmonic oscillators (\ref{Ham2}). Then, the 
$\star$-genvalue equation, reduces to a couple of simple ordinary differential equations
\beq
\left(\frac{z_{i}}{4}-z_{i}\partial^{2}_{z_{i}}-\partial_{z_{i}}-\frac{E}{2\hbar\Omega_{i}}\right)\rho(z_{1},z_{2})=0 \,, 
\hspace{5ex} i=1,2\,. 
\eeq
\noindent This equation acquires the same 
form as that for the simple harmonic oscillator \cite{ZFC}, therefore, one may be easily 
convinced that the Wigner function is determined by
\beq\label{WFO}
\rho_{nm}(X_{1},P_{1},X_{2},P_{2})=\frac{(-1)^{n+m}}{\pi^{2}\hbar^{2}}
e^{-\frac{2H_{1}^{\mathrm{osc}}}{\hbar\Omega_{1}}}
e^{-\frac{2H_{2}^{\mathrm{osc}}}{\hbar\Omega_{2}}}
L_{n}\left(\frac{4H_{1}^{\mathrm{osc}}}{\hbar\Omega_{1}}\right)
L_{m}\left(\frac{4H_{2}^{\mathrm{osc}}}{\hbar\Omega_{2}}\right) \,,
\eeq
 where the energy spectrum $E$ results equal to the energy spectrum of the Pais--Uhlenbeck oscillator (\ref{genvalues}).

 We can obtain the wave function $\psi(X_{1},X_{2})$ in the 
coordinates $(X_1,X_2)$ by Fourier transforming 
in the momentum variables the Wigner function~(\ref{WFO}) adapted to our system, 
obtaining~\cite{Cohen}
\beq
\psi(X_1,X_2)=\frac{1}{\psi^{*}(0,0)}\int dP_1
dP_2\,\rho\left( \frac{X_1}{2},P_1,\frac{X_2}{2},P_2\right)e^{i P_1 X_1+P_2 X_2/\hbar} \,, 
\eeq
\noindent where the constant $\psi^{*}(0,0)$ 
may be determined up to a phase by normalization of $\psi(X_1,X_2)$. By using the following identity between Laguerre and Hermite polynomials, denoted by $H$'s~\cite{Gradshteyn}, 
\beq
\int_{-\infty}^{\infty}dx\left[ H_{n}(x-a)H_{n}(x+a)e^{-x^{2}}e^{-2i bx}\right]=2^{n}\sqrt{\pi}n!e^{-b^{2}}L_{n}\left( 2(a^{2}+b^{2})\right) \,,  
\eeq
\noindent the integral gives the wave function in the position space
\beq
\psi_{nm}(X_{1},X_{2})=\frac{(-1)^{n+m}}{\pi^{2}\hbar^{2}}e^{-\frac{\Omega_{1}^{2}X_{1}^{2}}{2}}e^{-\frac{\Omega_{2}^{2}X_{2}^{2}}{2}}H_{n}(\sqrt{\Omega_{1}}X_{1})H_{m}(\sqrt{\Omega_{2}}X_{2})\,.
\eeq

The final step consists in calculate the wave function associated to the Pais--Uhlenbeck oscillator. To this end, we make use of the quantum canonical transformation operator, defined in (\ref{QCT}), where the generator, $F(q,x,X_{1},X_{2})$, of the classical canonical transformation takes the specific form of (\ref{generator}). Explicitly, 
the wave function for the Pais--Uhlenbeck 
oscillator 
in the position space is given by the formula
\beq
\label{waveintegral}
\psi_{nm}(q,x)
& = & 
N\int\!\!\! dX_{1}dX_{2}\,
\left\{\exp\left[\frac{i}{\hbar}\left(\Omega_{1}\gamma qX_{2}+\gamma xX_{1}-\Omega_{1}^{2}qx-\Omega_{1}X_{1}X_{2} \right) \right]
\right.
\nn\\ 
& & 
\left.
\psi_{nm}(X_{1},X_{2})
\right\} \,, 
\eeq  
\noindent where $N$ is a normalization constant. From now on, we will consider $\hbar=1$
for simplicity.  Also, here we only state our results, 
leaving 
the technical 
details on the construction of the wave equation 
to the Appendix. Bearing this in mind, after performing the integration
in~(\ref{waveintegral}), we are able to determine the 
wave function for the Pais--Uhlenbeck oscillator, which explicitly
reads
\beq\label{wavefunction}
\psi_{nm}(q,x)=N_{nm}\exp\left[ -i\Omega_{1}\Omega_{2}qx\right] \exp\left[ -\frac{\Delta}{2}\left(x^{2}+\Omega_{1}\Omega_{2}q^{2} \right) \right]\phi_{nm}(q,x), 
\eeq
\noindent where $\Delta:=\Omega_{1}-\Omega_{2}$, and the functions $\phi_{nm}$ stand for
\begin{eqnarray}\label{phi}
\phi_{nm}(q,x)
& = & 
\sum_{k=0}^{m}A_\Delta^{k}\frac{m!(n-m)!}{(m-k)!k!(n-m+k)!}H^{+}_{n-m+k}H_{k}^{-} \,, \quad m\leq n \,, \nn\\
\phi_{nm}(q,x)
& = & \sum_{k=0}^{n}A_\Delta^{k}\frac{n!(m-n)!}{(n-k)!k!(m-n+k)!}H^{+}_{k}H_{m-n+k}^{-} \,, \quad m>n \,.
\end{eqnarray} 
\noindent Here, the constant $A_\Delta:=i\Delta/4\sqrt{\Omega_{1}\Omega_{2}}$ only depends on the frequencies, while the
$H^{+}$'s and the $H^{-}$'s represent 
Hermite polynomial evaluated in the arguments 
\beq
H^{+}_n & := & H_n\left[i\sqrt{\Omega_{1}}\left(\Omega_{2}q-i x \right)  \right] \,,\nn\\ 
H^{-}_n & := & H_n\left[i\sqrt{\Omega_{2}}\left(\Omega_{q}q+i x \right)  \right] \,,
\eeq 
respectively, and the constants $N_{nm}$ behave as normalization factors.
\noindent The wave function (\ref{wavefunction}) 
stand for the solution to the Schr\"odinger equation associated to the Hamiltonian operator of the Pais--Uhlenbeck oscillator \cite{Smilga}. As it was proven in~\cite{Ghost}, these solutions are normalized resulting in a pure point spectrum, and all eigenfunctions form a complete orthogonal basis in the Hilbert space $\L^{2}(\mathbb{R}^{2})$.


\subsection{Equal frequency limit}

In the equal frequency limit, 
$\Delta =\Omega_1-\Omega_2 \rightarrow 0$, the quantum wave function (\ref{wavefunction}) cease to be normalizable, implying that the spectrum acquires continuous values. Indeed, 
within our context the Wigner function (\ref{WF}) oscillates wildly but eventually approximating to zero, therefore, the probability distribution appears as a generalized function, in such a way that the energy of the system is equally likely to be found anywhere in an interval 
$\left[E,E+dE\right]$ for any 
$E$~\cite{Wave,Generating}. Explicitly, 
whenever $\Delta=0$ the source of the continuous spectrum can be traced out through a quantum canonical transformation (\ref{QCT}) defined via the linear classical canonical transformation
\beq
F(q,x,Q_{1},Q_{2})=\frac{qQ_{2}}{\sqrt{2}}-\frac{\Omega qx}{4}+\frac{\Omega xQ_{1}}{\sqrt{2}}-\frac{Q_{1}Q_{2}}{2} \,,
\eeq  
where we have considered the single frequency 
as $\Omega:=\Omega_{1}=\Omega_{2}$.
\noindent This generating function maps 
maps the Pais--Uhlenbeck Hamiltonian (\ref{Ham})
to a new Hamiltonian
\beq
\label{eq:HamOmega}
H_{\mathrm{\Omega}}=\Omega(Q_{1}P_{2}-Q_{2}P_{1})-\frac{\Omega^{2}}{4}(Q_{1}^{2}+Q_{2}^{2}) \,, 
\eeq
which clearly differs from that obtained 
as the difference of two uncoupled harmonic oscillators
in~(\ref{Ham2}).  We also may easily deduce that 
the spectrum of the equal frequency 
Hamiltonian~(\ref{eq:HamOmega}) is composed of a discrete spectrum coming from the angular momentum part $Q_{1}P_{2}-Q_{2}P_{1}$, and 
of a continuous spectrum originated from the squared norm of the position variables 
$Q_{1}^{2}+Q_{2}^{2}$ as demonstrated in Ref.~\cite{Kosinski}, 
\beq
E_{mk}=\Omega\hbar\left(m-\frac{\Omega\hbar k^{2}}{4} \right)  \,. 
\eeq

Moreover, as it was indicated in 
\cite{Smilga,Kosinski}, the Pais--Uhlenbeck system 
results ghost-free, as even in the equal frequency limit 
$\Delta\rightarrow 0$ the evolution operator is certainly unitary. 
Again, within our context, the 
unitary property is readily obtained from the Wigner functions (\ref{WF}) and (\ref{WFO}), as 
both are related by a quantum canonical transformation of the form (\ref{transobs}), where the unitary operator $\widehat{U_{T}}$ is in fact given by the integral transformation~(\ref{waveintegral}). Following this reasoning, since the time evolution of the pair of harmonic oscillators is a solution of Moyal's equation~\cite{ZFC}
\beq
\rho(X_{1},X_{2},P_{1},P_{2},t)=U_{\star}^{-1}\star\rho(X_{1},X_{2},P_{1},P_{2},0)\star U_{\star} \,,
\eeq     
\noindent where the $\star$-evolution operator 
\begin{eqnarray}
U_{\star}(X_{1},X_{2},P_{1},P_{2})&=&e^{i tH/\hbar}_{\star}:=1+\left(\frac{i t}{\hbar}\right)H+\frac{1}{2!}\left(\frac{i t}{\hbar}\right)^{2}H\star H+\cdots  \nn\\
&=&2\pi\hbar\sum_{n,m}e^{it E_{nm}/\hbar}\rho_{nm}  
\end{eqnarray}
\noindent corresponds to a unitary operator.
Therefore,  the evolution of the Pais--Uhlenbeck oscillator is also unitary, as a consequence.

\section{Discussion}
\label{sec:discussion}

In this article, we analyzed the 
fourth 
order Pais--Uhlenbeck model within the 
Deformation quantization formalism.
We obtained both, the Wigner distribution function and the 
wave equation, for this system.  In particular, in order to 
obtain the Wigner function we explicitly 
consider the symmetries of the system associated to its 
Noether charges  which were established as 
ad hoc variables for solving the $\star$-genvalue equation.  
For the wave equation, we proceed by first considering 
a classical canonical transformation that map the Pais--Uhlenbeck 
model to a system of uncoupled harmonic oscillators.  
Afterward, our strategy was to obtain the Wigner function 
for the new system, and then to construct the associated 
wave function 
for this system.  Finally, we reached the wave function 
for the Pais--Uhlenbeck model by a quantum canonical 
transformation.  This wave equation resulted identical 
to the one considered in~\cite{Smilga}. 
We also showed that the model contains a continuous 
spectrum and results ghost-free, the two conditions together
being a consequence of the unitariness of the relevant quantum canonical transformations.  

Further studies are necessary in order to see the generality 
of conserved quantities associated to Noether charges as appropriate variables to find solutions 
to the $\star$-genvalue equation for a generic model.  
This will be done elsewhere.

\section*{Acknowledgments}
JB--M acknowledges support from PROMEP-UASLP-PTC-477 and CONACyT. 
AM acknowledges financial support from Conacyt-Mexico project CB-2014-243433. ER acknowledges partial support from the PRODEP grant UV-CA-320: \'Algebra, Geometr\'ia y Gravitaci\'on, and from CONACyT grant CB-2012-01-177519-F.

\appendix
\section{Technical steps for the construction of the Pais--Uhlenbeck wave function}
\label{sec:technical}

In the following we will detail some calculations that were omitted in the main text. Explicitly, 
by means of the quantum canonical 
transformation~(\ref{waveintegral}), the wave function for the Pais--Uhlenbeck system (\ref{waveintegral}) can be expressed as
\begin{eqnarray}
\psi(q,x)&=&\exp\left[ -i\Omega_{1}^{2}qx\right] \int dX_{1}dX_{2}\exp\left[ i\gamma xX_{2}-\frac{\Omega_{2}X_{2}^{2}}{2}+i\Omega_{1}\gamma qX_{1}\right. \nn\\
&&\left. -i\Omega_{1}X_{1}X_{2}-\frac{\Omega_{1}X_{1}^{2}}{2}\right]H_{m}\left(\sqrt{\Omega_{1}}X_{1} \right)H_{m}\left(\sqrt{\Omega_{2}}X_{2} \right) \,.    
\end{eqnarray}
Defining $Z_{1}:=\sqrt{\Omega_{1}}X_{1}$ and $Z_{2}:=\sqrt{\Omega_{2}}X_{2}$, 
this expression may be written as \begin{eqnarray}
\psi(q,x)&=&\frac{\exp\left[-i\Omega_{1}^{2}qx \right] }{\sqrt{\Omega_{1}\Omega_{2}}}\int dZ_{1}dZ_{2}\exp\left[\frac{i\gamma xZ_{2}}{\sqrt{\Omega_{1}}}-\frac{Z_{2}^{2}}{2}+i\sqrt{\Omega_{1}}\gamma qZ_{1} \right. \nn \\
&&\left. -\sqrt{\frac{\Omega_{1}}{\Omega_{2}}}Z_{1}Z_{2}-\frac{Z_{1}^{2}}{2}\right]H_{n}(Z_{1})H_{m}(Z_{2}) \,.  
\end{eqnarray}
\noindent Multiplying both sides by $\sum_{n=0}^{\infty}\sum_{m=0}^{\infty}\frac{t^{n}}{n!}\frac{u^{m}}{m!}$, and using the generating function 
for the Hermite polynomials \cite{Gradshteyn}
\beq
\exp\left(-t^{2}+2tx \right)=\sum_{k=0}^{\infty}\frac{t^{k}}{k!}H_{k}(x)  \,,  
\eeq
\noindent we perform the $Z_{1}$ and $Z_{2}$ integration using the 
Gaussian integral \cite{Gradshteyn}
\beq
\int_{-\infty}^{\infty}dx\exp\left(-p^{2}x^{2}+qx \right)=\exp\left(\frac{q^{2}}{4p^{2}}\frac{\sqrt{\pi}}{p} \right)  \,,  
\eeq
thus obtaining in this way
 the identity 
\begin{eqnarray}
\label{eq:bigsum}
\sum_{k=0}^{\infty}\sum_{l=0}^{\infty}\frac{t^{k}}{k!}\frac{u^{k}}{l!}\psi(q,x)
&=&
2\pi\exp\left[ -i\Omega_{1}\Omega_{2}qx\right] \exp\left[ -\frac{\Delta}{2}\left(x^{2}+\Omega_{1}\Omega_{2}q^{2} \right) \right] \nn\\
&&\times 
\sum_{k=0}^{\infty}
\sum_{l=0}^{\infty}
\sum_{j=0}^{\infty}
\left\{
\left(-4i\sqrt{\Omega_{1}\Omega_{2}} \right)^{j}\left(\frac{(\Omega_{1}-\Omega_{2})^{(k+l)/2}}{(\Omega_{1}+\Omega_{2})^{(k+l+2j)/2}}
\right) 
\right.  \nn\\
& & \times 
\left.
\left(
\frac{t^{k+j}u^{l+j}}{k!l!j!}
\right)
H_{k}(x^{+})H_{l}(x^{-})
\right\}  \,.
\end{eqnarray} 
\noindent Using the double summation identities \cite{Schwatt},
\beq
\sum_{n=0}^{\infty}\sum_{k=0}^{\infty}A_{k,n}
& = & \sum_{n=0}^{\infty}\sum_{k=0}^{n}A_{k,n-k} \,,\nn\\
\sum_{k=0}^{\infty}\sum_{n=k}^{\infty}A_{k,n}
& = & 
\sum_{n=0}^{\infty}\sum_{k=0}^{n}A_{k,n}  \,, 
\eeq
\noindent and relabeling the summation indices, we 
finally obtain that the $n,m$-term of the 
sum~(\ref{eq:bigsum}), defined as $\psi_{nm}$,  
results identical to the expression~(\ref{wavefunction}).

\end{document}